# Mpemba paradox: Hydrogen bond memory and water-skin supersolidity

Chang Q Sun, ecqsun@ntu.edu.sg

Nanyang Technological University

Abstract

Numerical reproduction of measurements, experimental evidence for skin supersolidity and hydrogen bond (O:H-O) memory clarified that Mpemba paradox integrates the heat "emission-conduction-dissipation" dynamics in the "source-path-drain" cycle system. O:H-O bond memory enables it emits energy at a rate initial storage dependence; water-skin supersolidity raises the local thermal diffusivity, favoring outward heat flow; Non-adiabatic "source-drain" interface enables heat dissipation. Contribution from convection, evaporation, frost, supercooling, solutes, are insignificant.

## Contents

## 1.1 Observations: hot water freezes faster than its cold

The Mpemba effect [1-5] is the assertion that hot water freezes quicker than its cold, even though it must pass through the same lower temperature on the way to freezing. Insets in Figure 1 show the measured initial-temperature $\theta_i$ dependence of the thermal relaxation $\theta(\theta_i,t)$ and the temperature difference between skin and bulk $\Delta\theta(\theta_i,t)$ profiles of liquid water under identical conditions (purity, volume, drain temperature, etc.) [6], which demonstrate the following:

1) Hot water freezes faster than its cold under the same conditions;
2) The liquid temperature $\theta$ drops exponentially with cooling time (*t*) until the transition of water into ice, with a relaxation time $\tau$ that drops as $\theta_i$ is increased;
3) Water skin is warmer than sites inside the liquid and the skin of hotter water is even warmer throughout the course of cooling.

The ability of hot water freezing faster than cold seems counter-intuitive as it would seem that hot water must first become cold water and therefore the time required for this will always delay its freezing relative to cold water. As it is infrequently observed and disobeys the Newton's Law (1701) of thermodynamics, this phenomenon was named as Mpemba paradox, after Erasto Mpemba, a 13-year old student at the Magamba Secondary School in eastern Tanzania. Erasto asserted it in 1963 during making ice cream. However, the reason why this occurrence remains mystery since 350 B.C. when Aristotle noted [1]: "The water has previously been warmed contributes to its freezing quickly: for so it cools sooner".

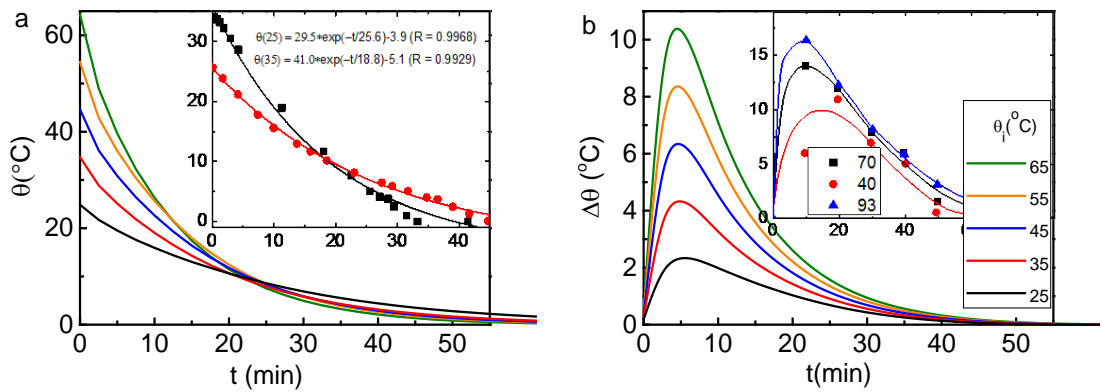

Figure 1. Numerical duplication [7] of the measured (insets) initial-temperature and time dependence of the (a) $\theta(\theta_i, t)$ [6] and the (d) skin-bulk temperature difference $\Delta\theta(\theta_i, t)$ profiles of water at cooling. Inset

(a) shows cooling and freezing of 30 ml deionised water at ≈ 25 and 35 °C in a glass beaker without cover or mixing using magnetic stirring. (Reprinted with permission from [7]).

*1.2    Explanation: integrated hydrogen-bond relaxation*

Numerical reproduction [8] of observations in Figure 1 and experimental evidence of hydrogen-bond (O:H-O) memory [7] and water-skin supersolidity [9] results in consistently the following mechanism:

1) Mpemba effect integrates the heat "emission-conduction-dissipation" dynamics in the "source-path-drain" cycle system [10].
2) O:H-O memory. Heating or skin molecular undercoordination (with fewer than four nearest neighbors in the bulk) stores energy into water by stretching the O:H nonbond and shortening the H-O bond (Figure 2a); cooling does the opposite to emit energy at a rate of history dependence (Figure 2b).
3) Water-skin is in supersolid state that is elastic, hydrophobic, less dense, and thermally more stable [9]. Skin supersolidity and heating lower the local mass density, which elevate the local thermal diffusivity, favoring outward heat flow.
4) Being sensitive to the source volume, skin radiation, and drain temperature, the Mpemba effect proceeds only at the strictly non-adiabatic 'source–drain' interface.

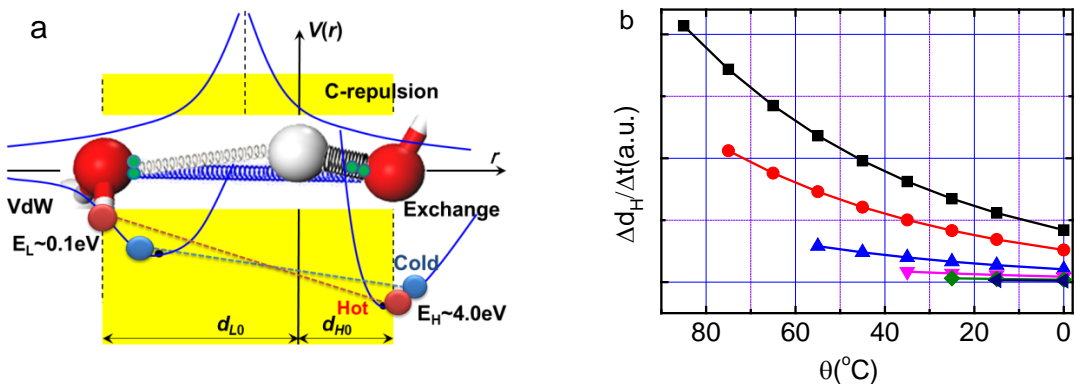

Figure 2. (a) O:H-O bond asymmetric, short-range potentials, being analogous to springs. Potentials include the O:H nonbond ($E_L$ ~0.1 eV cohesive energy; left-hand side), the H-O bond exchange interaction ($E_H$ ~ 4.0 eV; right-hand side), and the Coulomb repulsion between electron pairs (paired green dots) on adjacent oxygen atoms [11]. These interactions dislocate O atoms in the same direction by different amounts with respect to the $H^+$ origin under excitation. O ions dislocate along the O:H-O bond potentials from hotter (red line linked spheres, labeled 'hot') to colder state (blue line linked spheres,

labeled 'cold'). (b) O:H-O bond memory - initial temperature dependence of the H-O bond relaxation velocity at cooling, derived from experimental data ρ(θ) and θ(t) without any approximation or assumption. (Reprinted with permission from [7]).

Generally, heating stores energy into a 'normal' substance by stretching all bonds involved. However, heating lengthens the O:H part and shortens the H-O bond through O-O repulsion. Molecular undercoordination shortens the H-O bond and lengthen the O:H. The O:H elongates more than H-O contracts, resulting in ¼ density loss. The thermal diffusivity is inversely proportional to the density. Therefore, the skin thermal diffusivity is at least 4/3 greater than the bulk. As illustrated in Figure 3, $H_2O$ molecules become smaller and their separation larger when water is heated or at the skin. When the water is exposed to cool environment, O:H-O bond relaxation occurs, which analogues the process of releasing a coupled pongee pair whose velocity depends on its initial deformation or energy storage.

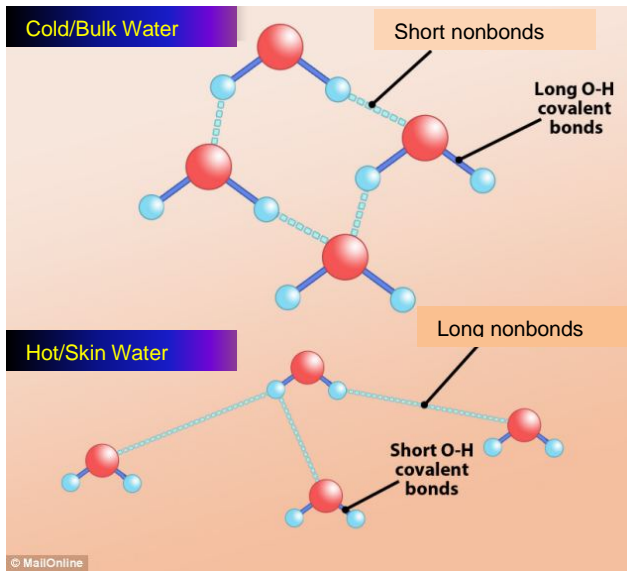

Figure 3 O:H-O bond segmental length in hot/skin water compared with that in cold/bulk water. Heating or molecular undercoordination shrinks molecular size but enlarges the separation between molecules, with density loss. Water has memory to reverse the cooling/heating in energy gain and loss [12].

Figure 4a shows the initial temperature $\theta_i$ dependence of cooling and freezing duration $t_i$ [2] and the relaxation time $\tau_i$ [7] of energy loss. The $t_i$ and the $\tau_i$ drops almost exponentially with the increase of the $\theta_i$ (Figure 4a). For instance, making ice using 78 °C water takes 40 minutes with a 15 minutes relaxation time; making ice from 18 °C water takes 100 minutes with a relaxation time of 75 minutes. The $\tau_i$ also drops with the increase of initial energy storage, or the initial vibration frequency [13], as Figure 4b shows.

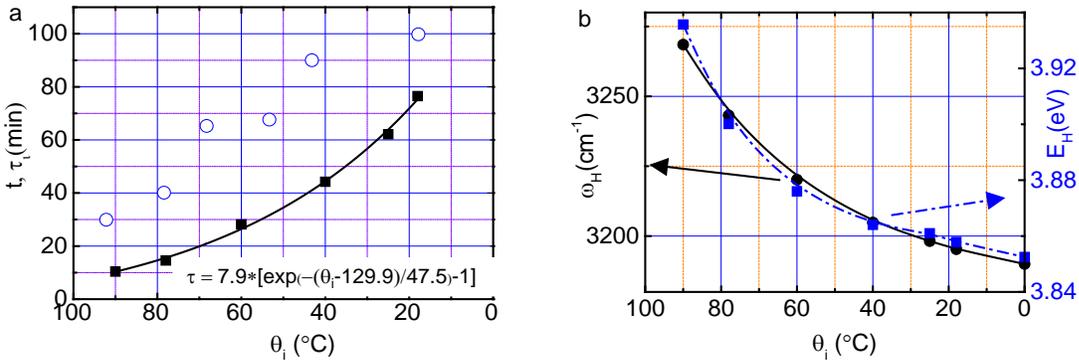

Figure 4. Initial temperature $\theta_i$ dependence of (a) the cooling time $t$ (scattered circles), relaxation time $\tau_i$ (fitted solid line), and (b) the measured initial energy $E_H$ (solid black line) and vibration frequency $\omega_H$ (broken blue line) for liquid water [13]. (Reprinted with permission from [7]).

*1.3 Indication*

The H-O bond energy is 3.97 eV for bulk water. For the undercoordinated monolayer it is 4.66 and for vapor monomer molecules, it is 5.10 eV. The present understanding of the Mpemba paradox indicates typically the following:

1) **Hydrogen generation**. Splitting the H-O bond in cold bulk water requiring 3.97 eV per bond would be easier than in the hot or undercoordinated molecules (> 3.97 eV because of cooling contraction) in the skin (4.66 eV) or vapor (5.10 eV), irrespective of the catalyst used.
2) **Ice cube fabrication**. Hotter water or water under the sun freezes faster.
3) **Snow production.** Making snow (~ 4.66 eV) using undercoordinated hot vapor (> 5.10 eV) needs less energy than making ice (3.97 eV) from the perspective of the H-O bond energy. A snapshot of a video clip in Figure 5 shows the making of snow from spreading hot water at -25 °C temperature [14].

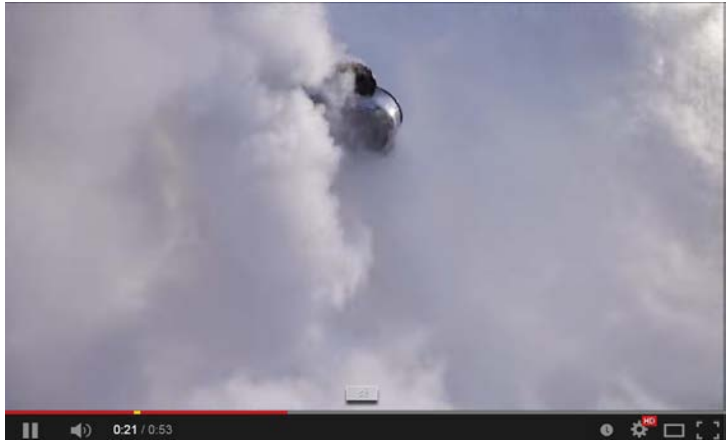

Figure 5 Turning boiling or hot water into snow at -25°C (-13°F)[14].

*1.4    History*

The phenomenon by which, under certain conditions, hot water freezes faster than cold water has been observed by many of the world's finest minds extending from 4[th] BC to date.

**Aristotle** (384–322 BCE) was a Greek philosopher and scientist. He observed in 350 BC, that "The fact that the water has previously been warmed contributes to it's freezing quickly: for so it cools sooner. Hence many people, when they want to cool water quickly, begin by putting it in the sun".

**Giovanni Marliani** (1420-1483) was an Italian physicist, doctor, philosopher and astrologer. He was the first to empirically prove that hot water freezes faster than its cold in 1461. He used four ounces of unheated water and four ounces of boiled water, which he placed in similar containers outside on a cold winter's day. He eventually observed that the boiled water froze first; although he was unable to explain the mechanics of how it happened.

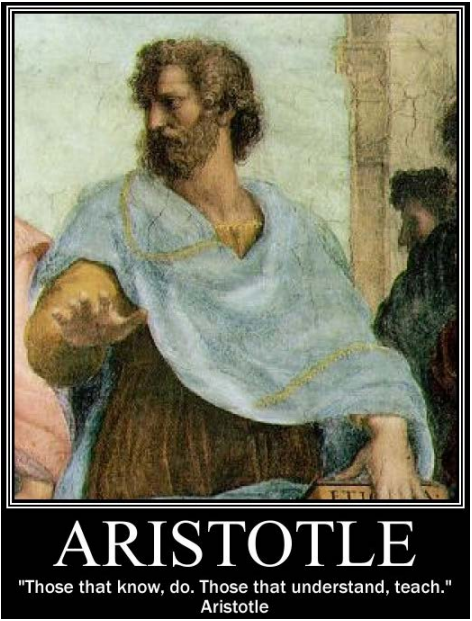

*Aristotle (384 -322 BC)* firstly found the Mpemba effect in 350 BC.

**Francis Bacon** (1561-1626) an English philosopher, statesman, scientist, jurist, orator, essayist, and author, wrote that "*slightly tepid water freezes more easily than that which is utterly cold*".

**Rene Descartes** (1596 –1650), a French philosopher, mathematician and writer, also tried to solve the problem in 1637 and throughout his years. He wrote in his *Discourse on the Method*, "*One can see by experience that water that has been kept on a fire for a long time freezes faster than other, the reason being that those of its particles that are least able to stop bending evaporate while the water is being heated.*"[5]

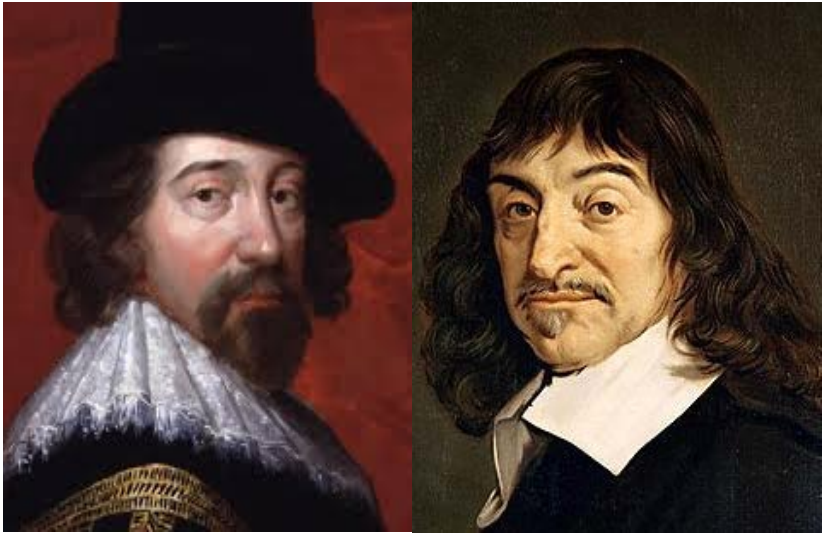

Francis Bacon (1561-1626); René Descartes (1596-1660)

**Erasto Mpemba** (1950-,Tanzanian) and **Denis Osborne** (1932-2014, Physicist and Educator). Mpembacatapulted back this question into the public eye in 1963 when he was a 13-year old boy. He noticed that if he put hot ice-cream mix into the freezer, it froze more quickly than the cooled ice-cream mix of his fellow classmates. Mpemba's tenacity in the face of his teacher's initial dismissal of his observations and the ridicule of his classmates prompted him to repeat the experiment with hot and cold water, and to stand up and ask visiting physics lecturer Denis Osborne about the phenomenon. Mpemba and Osborne published a paper together [2] in 1969, and the effect became known as *The Mpemba Effect*. This paper was republished by IoP in 1979.

Mpemba and Osborne placed 70 ml samples of water in 100 ml beakers in the ice box of a domestic refrigerator on a sheet of polystyrene foam. They recorded the time for *freezing to start* was longest with an initial temperature of 25 °C and that it was much less at around 90 °C. They ruled out loss of liquid volume by *evaporation* as a significant factor and the effect of *dissolved air*. In their setup most heat loss was found to be from the *liquid skin* [2].

In August 2012 Osborne described the following about his work with Mpemba: "In line with his question made in front of his school staff and peers, we tested and found that hot water in Pyrex beakers on polystyrene foam in a domestic freezer froze before cooler samples. We attributed this to *convection* creating a continuing hot top, noting that:
1) If two systems are cooled, the water that starts hotter may freeze first, but we did not look for ice and measured the time as that until a thermocouple in the water read 0°C.

2) A graph of 'time to start freezing' against initial temperature showed that the water starting at about 26°C took longest to freeze (water starting at 60°C took twice as long as water starting at 90°C).
3) Thermocouples near top and bottom showed a temperature gradient in the water. A hot starter kept a hot top while its lower levels were cooler than for the cool starter.
4) An oil film on the water surface delayed freezing for several hours, suggesting that without this film, most of the heat escaped from the top surface.
5) Changes in volume due to evaporation were small; the latent heat of vaporization for all the water to cool to 0°C and start freezing accounted for less than 30% of the cooling.
6) We used recently boiled water for all the trials, making dissolved air an unlikely factor. We failed to check and report the ambient temperature in the freezer or its consistency during cooling. Lower ambient air temperatures might increase heat loss rates from the top surface, cause more rapid convection and increase the difference in freezing times.
7) Different mechanisms may assume more importance in different situations. We gave one example, with Mpemba's initial discovery in mind, and we wrote: 'rapid cooling of a system that starts hot may be accelerated if it establishes thermal contact with the case of the freezer cabinet through melting the layer of ice and frost on which it rests'."

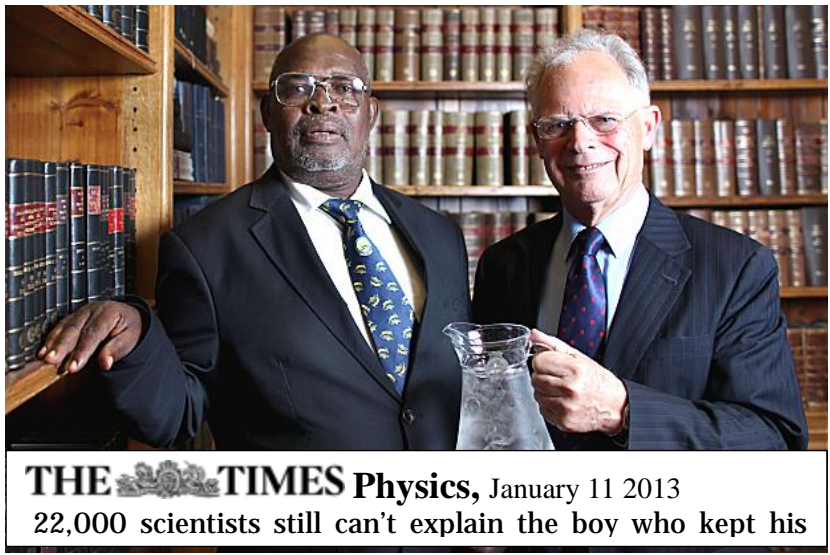

Erasto Mpemba (left) and Professor Denis Osborne (right) who jointly published [2] presented at the presentation in London (Courtesy of Ben Gurr/The Times).

David Auerbach (Writer, software engineer, critic) [15] described how the effect can be observed in samples in glass beakers placed into a liquid cooling bath. In all cases the water supercools, reaching a temperature of typically -6 °C to -18 °C before spontaneously freezing. Considerable random variation was observed in the time required for spontaneous freezing to start and in some cases this resulted in the water which started off hotter (partially) freezing first.

**James Brownridge** (1937-, a radiation safety officer at the State University of New York) [16, 17] had spent 10 years and conducted more than 20 experiments to examine all possible factors. He suggested that *supercooling* is the dominant factor.

**Philip Ball** (1962-, an English scientific writer)[18], noted in *Physics World*, "Even if the Mpemba effect is real - it is not clear whether the explanation would be trivial or illuminating." He pointed out that investigations of the phenomenon need to control a large number of initial parameters (including type and initial temperature of the water, dissolved gas and other impurities, and size, shape and material of the container, and temperature of the refrigerator) and need to settle on a particular method of establishing the time of freezing, all of which might affect the presence or absence of the Mpemba effect. The required vast multidimensional array of experiments might explain why the effect is not yet understood.

In 2012, the Royal Society of Chemistry held a competition calling for papers offering explanations to the Mpemba effect. More than 22,000 people entered and Erasto Mpemba himself announced **Nikola Bregović** [6] as the winner suggesting that *convection* and *supercooling* were the reasons for the effect.

Nikola Bregovic who works at the Laboratory of Physical Chemistry in the Department of Chemistry of the University of Zagreb, explained in the following:

"*The statement by Brownridge, 'Hot water will freeze before cooler water only when the cooler water supercools, and then, only if the nucleation temperature of the cooler water is several degrees lower than that of the hot water. Heating water may lower, raise or not change the spontaneous freezing temperature,' summarizes in great part the conclusions that may be drawn from almost all the data I have collected myself and others presented earlier. However, the effect of convection, which enhances the probability of warmer water freezing should be emphasized in order to express a more complete explanation of the effect. The fact that this effect is not fully resolved to this day, was an indication to me that fundamental problems lie underneath it, but still I did not expect to find that water could behave in such a different manner under so similar conditions. Once again this small, simple molecule amazes and intrigues us with it's magic*."

**Chang Qing Sun** (1956-, this author) and coworkers started to investigate this effect by examining the process at a molecular level from the perspective of O:H-O bond relaxation [19]. They deposited their preliminary work in ArXiv Physics [8] in October 2012. This explanation attracted much public attention and was featured by numerous media like *The Times, ArXiv Media, The Telegraph, Daily Mails, Chemical World, Chem Views, Physics Today, IOP Physics, Nature Chemistry, etc.*

This group of researchers experimentally verified the effects of O:H-O bond memory and water skin supersolidity, numerically reproduced observations of Mpemba and **Bregović** by solving the Fourier thermo-fluid transportation dynamics using finite element method with involvement of skin effect and appropriate boundary conditions [7]. They suggested an integrated mechanism that the Mpemba effect integrates the heat "emission-conduction-dissipation" dynamics proceeds in the "source-path-drain" cycle system. The steps of heat emission and conduction are intrinsically related to the O:H-O bond thermal and undercoordination relaxation dynamics and the passivation is related to the extrinsic operation conditions [7, 11].

### 1.5 Notes on existing explanations

The hypothetic factors explaining this effect include:
1) *Evaporation* [20]: The evaporation of the warmer water reduces the mass of the water to be frozen. Evaporation is endothermic, meaning that the water mass is cooled by vapor carrying away the heat, but this alone probably does not account for the entirety of the effect. Experiments confirmed that the mass loss is only 1.5% or less when cooled from 75 to -40 °C [7].
2) *Convection* [21-23]: Accelerating heat transfers. Reduction of water density below 4 °C (39 °F) tends to suppress the convection currents that cool the lower part of the liquid mass; the lower density of hot water would reduce this effect, perhaps sustaining the more rapid initial cooling. Higher convection in the warmer water may also spread ice crystals around faster. Numerical examination revealed however, that contribution of the convection velocity to the intersecting temperatures of two $\theta(\theta_i,t)$ relaxation curves is insignificant [7].
3) *Frost* [4, 17]: Frost insulates thermal dissipation. The lower temperature water will tend to freeze from the top, reducing further heat loss by radiation and convection, while the warmer water will tend to freeze from the bottom and sides because of water convection. This is disputed as there are experiments that account for this factor.

4) *Supercooling* [17, 23, 24]: When placed in a freezing environment, cool water supercools more than hot water in the same environment, thus solidifying slower than hot water. However, super-cooling tends to be less significant where there are particles that act as nuclei for ice crystals, thus precipitating rapid freezing. Experimentally derived crossing temperatures at θ > 0 °C (Figure 1a) involves no supercooling effect.

5) *Solutes* [25]: The effects of calcium carbonate, magnesium carbonate among others.[http://www.wikiwand.com/en/Mpemba_effect - citenote13](http://www.wikiwand.com/en/Mpemba_effect - citenote13) The Mpemba effect can be observed in deionized water without any solutes being involved.

6) *Thermal conductivity*: The container of hotter liquid may melt through a layer of frost that is acting as an insulator under the container (frost is an insulator), allowing the container to come into direct contact with a much colder lower layer that the frost formed on (ice, refrigeration coils, etc.) The container now rests on a much colder surface (or one better at removing heat, such as refrigeration coils) than the originally colder water, and so cools far faster from this point on.

7) *Dissolved Gases* [4]: Cold water can contain more dissolved gases than hot water, which may somehow change the properties of the water with respect to convection currents, a proposition that has some experimental support but no theoretical explanation.

However, all above factors are extrinsic to the energy 'emission–conduction–dissipation' dynamics in the 'source–path–drain' cycle system in which the Mpemba paradox takes place. Focusing on the nature and relaxation dynamics of the O:H-O bond [26] that is the primary constituent of the liquid source and path in this event.

1.6    Numerical justification: Water-skin supersolidity

1.6.1    Fourier thermal–fluid transport dynamics

Figure 6 illustrates an adiabatically-walled, open-ended, one-dimensional tube cell for solving the thermal fluid transport problem using the finite element method [27]. Water at an initial temperature $\theta_i$ in the cell is divided along the *x*-axis into two regions: the bulk (B, from $-L_1$ = -9 mm to 0) and the skin (S, from 0 to $L_2$ = 1 mm). The tube is cooled in a drain of constant temperature $\theta_f$ that is subject to variation to allow it to be examined for sensitivity.

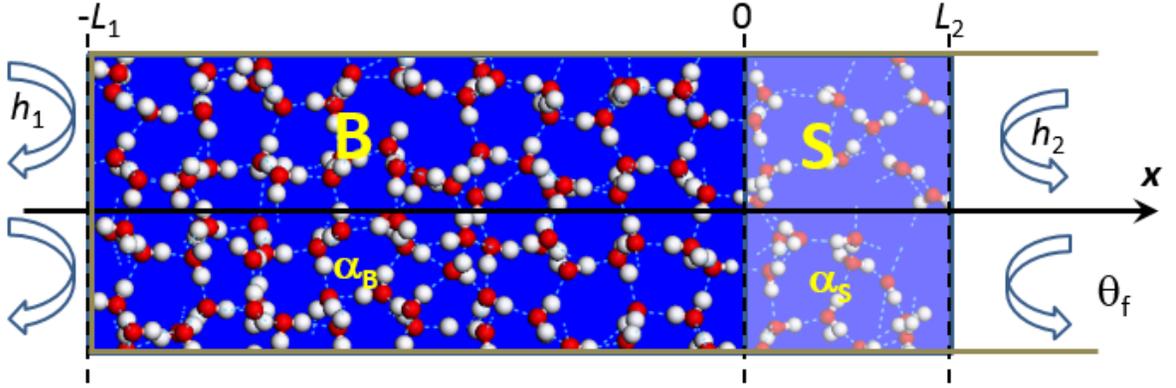

Figure 6. The water in the adiabatically walled, open-ended, one-dimensional tube cell at initial temperature $\theta_i$ is cooled in the drain of constant temperature $\theta_f$. The liquid source is divided along the x-axis into the bulk $B$ ($-L_1 = -9$ mm, 0) and the skin $S$ (0, $L_2 = 1$ mm), with respective thermal diffusivities $\alpha_B$ and $\alpha_S$. The mass densities are $\rho_S/\rho_B = 3/4$ [28, 29] in the respective region. The bulk-skin interface is at $x = 0$; $h_j$ is the heat transfer (radiation) coefficient at the tube ends, with the absence ($j = 1$, left-hand end) and presence ($j = 2$, right-hand end) of the supersolid skin.(Reprinted with permission from [7]).

The time-dependent gradient of temperature changes at any site ($x$), which follows the following function and the initial- and boundary conditions:

$$\begin{cases} \dfrac{\partial \theta(x)}{\partial t} = \nabla \cdot (\alpha(\theta(x), x) \nabla \theta(x)) - v \cdot \nabla \theta(x) \\ \alpha(\theta, x) = \dfrac{\kappa_B(\theta, x)}{\rho_B(\theta, x) C_{pB}(\theta, x)} \times \begin{cases} 1 & (Bulk) \\ \approx \rho_B/\rho_S (= 4/3) & (Skin) \end{cases} \\ v_S = v_B = 10^{-4} (m/s) \end{cases},$$

$$\begin{cases} \theta = \theta_i & (t = 0) \\ \theta(0^-) = \theta(0^+); \theta_x(0^-) = \theta_x(0^+) & (x = 0) \\ h_i(\theta_f - \theta) \pm \kappa_i \theta_x = 0 & (x = -l_1; l_2) \end{cases}.$$

(1)

The first term describes thermal diffusion and the second term describes thermal convection in the Fourier transport equation, where $\alpha$ is the thermal diffusivity and $v$ is the convection rate. The known temperature-dependence of the thermal conductivity $\kappa(\theta)$, the mass density $\rho(\theta)$, and the specific heat under constant pressure $C_p(\theta)$, given in Fig A1-1, determines the thermal diffusivity $\alpha_B$ of the bulk water. The skin supersolidity [28] contributes to $\alpha_S$ in the form $\alpha_S(\theta) \approx 4/3\alpha_B(\theta)$, because the skin mass density 0.75 gcm$^{-3}$ is 3/4 times the standard density at 4°C. $\alpha_S(\theta)$ is subject to optimization as the skin supersolidity may modify the $\kappa(\theta)/C_p(\theta)$ value in a yet unknown manner.

The boundary conditions represent that the temperature $\theta$ and its gradient $\theta_x = \partial\theta/\partial x$ continue at the interface ($x = 0$) and the thermal flux $h(\theta_f - \theta)$ is conserved at each end for $t > 0$. The convection velocity $v$ takes the bulk value of $v_S = v_B = 10^{-4}$ m/s, or zero for examination. As the heat transfer (through radiation) coefficient $h_j$ depends linearly on the thermal conductivity $\kappa$ in each respective region [30], the standard value of $h_1/\kappa_B = h_2/\kappa_S = 30$ w/(m$^2$K) [31] is necessary for solving the problem. The $h_2/\kappa_S$ term contains heat reflection by the boundary. The ratio $h_2/h_1 > 1$ describes the possible effect of skin thermal radiation.

1.6.2    Roles of convection, diffusion, and boundaries

The computer reads in the digitized $\rho(\theta)$, $\kappa(\theta)$ and $C_p(\theta)$ to compose the $\alpha_B(\theta)$ before each iteration of calculating the partitioned elemental cells. Besides the thermal diffusivity and the convection velocity in the Fourier equation, systematic examination of all possible parameters in the boundary conditions, shown in Figure 7 and Figure 8, revealed the following:

1) Characterized by the intersecting temperature, the Mpemba effect happens only in the presence of the supersolid skin ($\alpha_S/\alpha_B > 1$).
2) Complementing skin supersolidity, thermal convection distinguishes only slightly difference temperature $\Delta\theta$ between the skin and the bulk, and raises negligibly the crossing temperature.
3) The Mpemba effect is sensitive to the source volume, the $\alpha_S/\alpha_B$ ratio, the radiation rate $h_2/h_1$ and the drain temperature $\theta_f$.
4) The bulk/skin thickness ($L_1:L_2$) ratio and the thermal convection have little effect on observations.

For instance, increasing the liquid volume may annihilate the Mpemba effect because of the non-adiabatic process of heat dissipation. It is understandable that cooling a drop of water (1 mL) needs shorter time than cooling one cup of water (200 mL) at the same $\theta_i$ and under the same conditions. Higher skin radiation $h_2/h_1 > 1$ promotes the Mpemba effect. Therefore, conditions for the Mpemba effect are indeed very critical, which explains why it is not frequently observed. Figure 1 shows the numerical reproduction of the observed Mpemba attributes (insets) [2, 6].

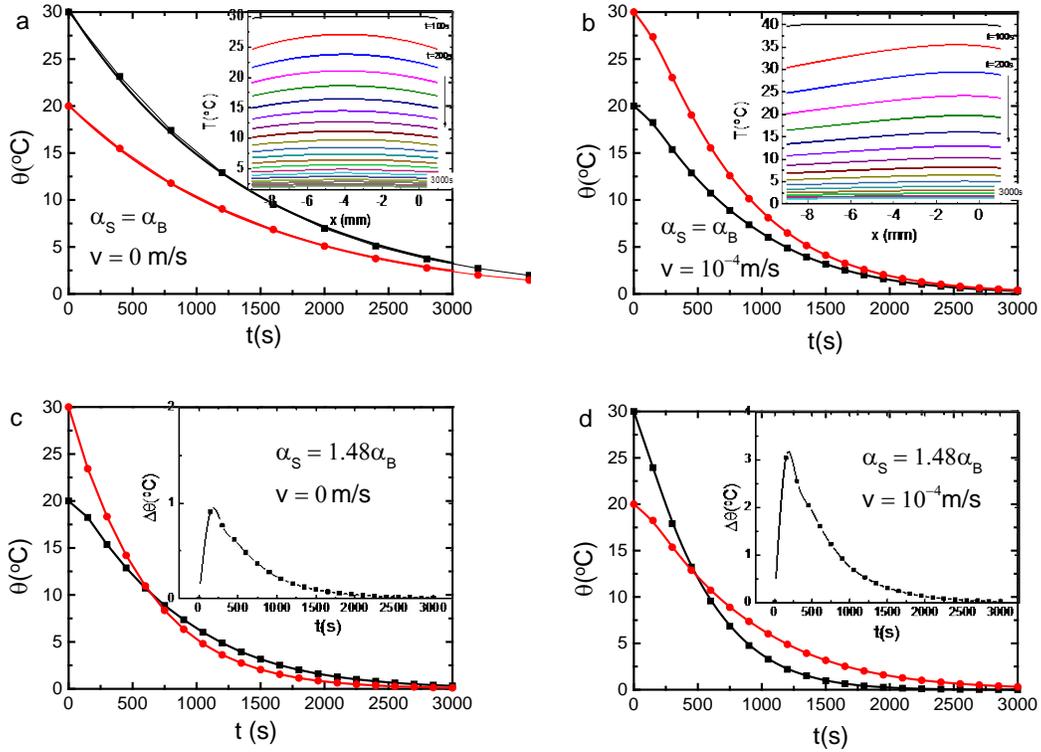

Figure 7. Thermal relaxation curves $\theta(\theta_i, t)$ at $x = 0$: (a), (b) with supersolid skin absent ($\alpha_S/\alpha_B = 1$); and (c), (d) with supersolid skin present (optimized at $\alpha_S/\alpha_B = 1.48$); and (a), (c) with thermal convection absent ($v_S = v_B = 0$); and (b), (d) with thermal convection present ($v_S = v_B = 10^{-4}$ m/s). The Mpemba effect is characterized by the crossing temperature which occurs only in the presence of the skin supersolidity, irrespective of the thermal convection. The insets in (a) and (b) show the time-dependent thermal field in the tube. Supplementing the skin supersolidity, convection only slightly raises $\Delta\theta$ and the crossing temperature — as the insets in (c) and (d) show. (Reprinted with permission from [7]).

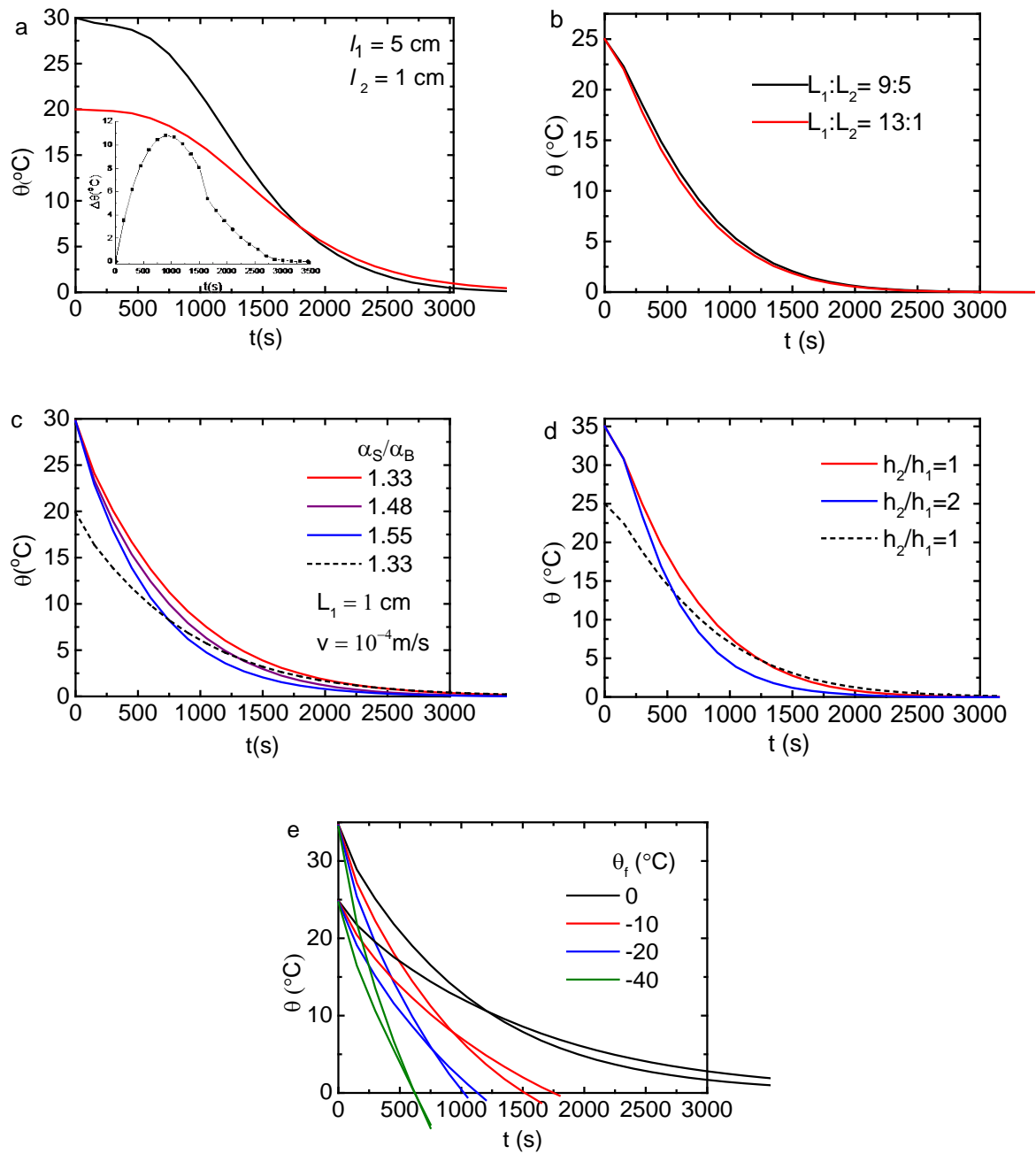

Figure 8. Sensitivity of the Mpemba effect (crossing temperature) to: (a) the source volume; (b) the bulk/skin thickness ratio ($L_1$:$L_2$); (c) the supersolidity ratio $\alpha_S/\alpha_B$; (d) the radiation rate $h_2/h_1$; and (e) the drain temperature $\theta_f$. Volume inflation (from 1 to 5 cm) in (a) prolongs the time until the crossing temperature is reached, and raises the skin temperature (see inset). (b) The $L_1$:$L_2$ ratio has little effect on the relaxation curve. Increasing (c), the $\alpha_S/\alpha_B$ and (d) the $h_2/h_1$ ratio promotes the Mpemba effect. (e) Lowering the $\theta_f$ shortens the time until the crossing temperature is reached. The sensitivity examination is conducted based on the conditions of $\alpha_S/\alpha_B = 1.48$, $v_S = v_B = 10^{-4}$ m/s, $\theta_f = 0°C$, $L_1 = 10$ mm, $L_2 = 1$ mm, $h_1/\kappa_B = h_2/\kappa_S = 30$ w/(m²K) unless indicated. (Reprinted with permission from [7]).

1.7    Experimental justification: O:H-O bond memory

1.7.1    O:H-O bond relaxation velocity

The following formulates the decay curve $\theta(\theta_i, t)$ shown in Figure 1a [6]:

$$\begin{cases} d\theta = -\tau_i^{-1}\theta dt & (decay\ function) \\ \tau_i^{-1} = \sum_j \tau_{ji}^{-1} & (relaxation\ time) \end{cases},$$

(2)

where the $\theta_i$-dependent relaxation time $\tau_i$ is the sum of $\tau_{ji}$ over all possible $j$-th process of heat loss during cooling.

A combination of the measured $\theta(\theta_i, t)$ (Figure 1a inset) and the $d_H(\theta)$ (Fig A1-1a converts into Fig A1-1b) reveals the memory of the O:H-O bond without needing any assumption or approximation. The $\theta(\theta_i, t)$ curve provides the slope of $d\theta/dt = -\tau_i^{-1}\theta$ and the $d_H(\theta) = 1.0042 - 2.7912 \times 10^{-5} \exp\left[(\theta+273)/57.2887\right]$ (Å) formulates the measured $\theta$ dependence of the H-O bond relaxation. Combining both slopes immediately yields the linear velocity of $d_H(\theta)$ relaxation at cooling.

As the O:H nonbond and the H-O bond are correlated, the relaxation velocities of their lengths and energies are readily available, since $E_x = k_x(\Delta d_x)^2/2$ approximates the energy storage with the known $d_H$ velocity. For simplicity and conciseness, the focused will be on the instantaneous velocity of $d_H$ during relaxation:

$$\frac{d(d_H(\theta))}{dt} = \frac{d(d_H(\theta))}{d\theta}\frac{d\theta}{dt} = -\tau_i^{-1}\theta\frac{\Delta(d_H(\theta))}{57.2887},$$

where:

$$\Delta(d_H(\theta)) = -2.7912 \times 10^{-5} \exp[(\theta+273)/57.2887].$$

(3)

Figure 2b plots the $\theta_i$-dependence of the $d_H$ linear velocity, which confirms that the O:H-O bond indeed possesses memory. Although passing through the same temperature on the way to freezing, the initially shorter H-O bond at higher temperature remains highly active compared to its behavior otherwise when they meet on the way to freezing.

1.7.2 Relaxation time versus initial energy storage

Solving the decay function Eq. (3) yields the relaxation time $\tau_i(t_i, \theta_i, \theta_f)$:

$$\tau_i = -t_i \left[ Ln\left(\frac{\theta_f + b_i}{\theta_i + b_i}\right) \right]^{-1}.$$

(4)

An offset of the $\theta_f$ (= 0°C) and the $\theta_i$ by a constant $b_i$ is necessary to ensure $\theta_f + b_i \geq 0$ in the solution ($b_i = 5$ was taken with reference to the fitted data in Figure 1a). A combination of the known $t_i$, $\theta_i$ and $\theta_f$, given in Figure 4a (scattered data) yields the respective $\tau_i$, shown by the solid line. The $\tau_i$ drops exponentially with the increase of the $\theta_i$ (Figure 4a), or with the increase of initial energy storage, or the initial vibration frequency measurements [13], as Figure 4b shows.

1.8 Further discussion: heat 'emission–conduction–dissipation' dynamics
1.8.1 Liquid source and path: Heat emission and conduction

Figure 2a illustrates the cooperative relaxation of the O:H-O bond in water under thermal cycling. An interplay of the O:H vdW-like force, the H-O exchange interaction, the O--O Coulomb repulsion, the specific-heat disparity between the O:H and the H-O bond, always dislocate O atoms in the same direction along the respective potential paths [32].

Generally, heating stores energy in a substance by stretching all bonds involved. However, heating excitation stores energy in water by lengthening the O:H nonbond. The O:H expansion weakens the Coulomb interaction, which shortens the H-O bond by shifting the $O^{2-}$ towards the $H^+$ (red line linked spheres in Figure 2a are in the hot state). Cooling does the opposite (blue line liked spheres), analogous to suddenly releasing a pair of coupled, highly deformed springs, one of which is stretched and the other compressed. This relaxation emits energy at a rate that depends on the deformation history (i.e., how much they were stretched or compressed). Energy storage and emission of the entire O:H-O bond occurs mainly through H-O relaxation, since $E_L$ (about 0.1 eV) is only 2.5% of $E_H$ (about 4.0 eV) [28]. The O:H-O bond memory and the unusual way of energy emission yield the history-dependent H-O bond relaxation velocity at cooling, as shown in Figure 2a. O:H-O bond cooperative relaxation under thermal cycling [33, 34]. The $d_{H0}$ and $d_{L0}$ are the respective references at 4°C. Indicated are the O:H vdW-like nonbond interaction ($E_L$ ~0.1 eV; left-hand side), the H-O bond exchange interaction ($E_H$ ~ 4.0 eV; right-hand side), and the inter-oxygen electron-pair repulsion (paired green dots). A combination of these interactions and the specific-heat disparity between the O:H and the H-O dislocate O atoms in the same direction by different amounts when cooling. The relaxation proceeds along the O:H-O bond potentials from hotter (red line linked spheres, labeled 'hot') to colder state (blue line linked spheres, labeled 'cold').

During liquid heating, molecular undercoordination has the same effect as above on O:H-O bond relaxation. Heating and molecular undercoordination are mutually enhanced during O:H-O bond relaxation and the associated thermal diffusivity in the skin region. Mass density is lowered, raising the thermal diffusivity, which favors outward heat flow in the conduction path.

1.8.2   Source–drain interface: Nonadiabatic cycling

Mpemba effect happens only in circumstances where the water temperature drops abruptly from $\theta_i$ to $\theta_f$ at the source–drain interface [7]. Examination has indicated that the Mpemba crossing temperature is sensitive to the volume of the liquid source (Figure 8a). Larger liquid volumes may prevent this effect by heat-dissipation hindering. As confirmed by Brownridge [17], any spatial temperature decay between the source and the drain could prevent the Mpemba effect. Examples of such decay include sealing the tube ends, an oil film covering, a vacuum isolating the source–drain chamber, muffin-tin-like containers connecting, or multiple sources contributing to a limited fridge volume. Conducting experiments under identical conditions is necessary to minimize artifacts such as radiation, source/drain volume ratio, exposure area, container material, etc.

Figure 9 shows the $\theta(\theta_i,t)$ profiles obtained under different conditions, showing that the O:H-O bond is very sensitive to stimulus during relaxation. Conditions for the Mpemba effect are indeed very critical, which explains why it is not frequently observed.

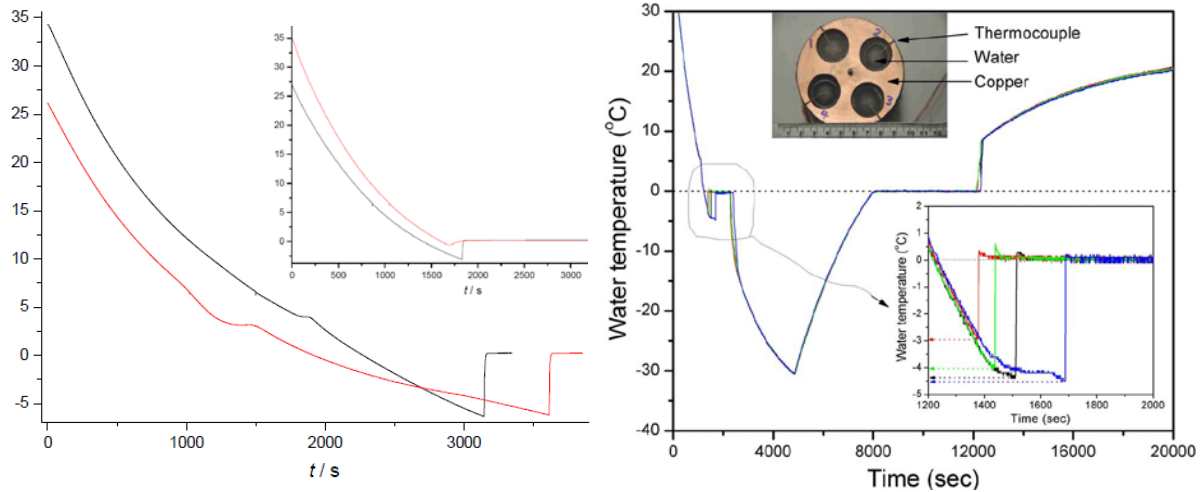

Figure 9 Cooling and freezing of 30 ml deionized water in icebox without stir mixing. The inset in beaker with magnetic stir mixing [6]. Four specimens of tap hard water in a copper container designed to maintain identical cooling conditions for each specimen [16].

1.8.3    Other factors: Supercooling and evaporating

Water molecules with fewer than four nearest neighbors, such as those that form the skin, a monolayer film, or a droplet on a hydrophobic surface or hydrophobically confined are subject to stretching dispersion of the quasi-solid phase boundaries. The phase boundary dispersion results in the melting point elevation and the freezing point depression. Approximately, the $E_H$ determines the critical temperature for melting and the $E_L$ dominates freezing [32]. The critical temperature elevation/depression is different from the process of superheating/superheating. Supercooling is associated with the initially longer O:H bond at higher temperature, which contributes positively to the liner velocity of O:H-O bond relaxation with a thermal momentum at cooling. Supercooling of colder water can never happen because of the relatively stiffer O:H nonbond with higher $\Theta_{DL}$ value. Therefore, the colder water reacts more slowly to the relaxation at freezing than the bonds in the warmer water, because of the lower momentum of relaxation – memory effect.

The involvement of ionic solutes or impurities [35, 36] mediates Coulomb coupling because of the alternation of charge quantities and ion volumes [37]. Salting shifts the H-O phonon positively in the same way as heating [38-40] to weaken the Coulomb repulsion. Salting or impurities are expected to raise the velocity of heat emission at cooling but in Mpemba's observation, the only variable is temperature. Mass loss due to evaporation [3] has no effect on the relaxation rate of the O:H-O bond. The mass loss for samples at slightly different temperature is negligible.

## 1.9 Summary

Reproduction of observations revealed the following:

1) Mpemba effect integrates the heat "emission-conduction-dissipation " dynamics in the "source-path-drain" cycle system as a whole. One cannot separate these processes and treat them independently.
2) O:H-O bonds possess memory whose thermal relaxation defines intrinsically the rate of energy emission. Heating stores energy in water by O:H-O bond deformation. The H-O bond is shorter and stiffer in hotter water than its cold. Cooling does the opposite, emitting energy with a thermal momentum that is history-dependent.
3) Heating enhances the skin supersolidity that elevates the skin thermal diffusivity with a critical ratio of $\alpha_S/\alpha_B \geq \rho_B/\rho_S = 4/3$. Convection alone produces no Mpemba effect.
4) Mpemba effect proceeds only in the highly nonadiabatic "source-drain" interface to ensure immediate energy dissipation. The Mpemba crossing temperature is sensitive to the volume of liquid source being cooled, the drain temperature and skin radiation.
5) The Mpemba effect takes place with a characteristic relaxation time that drops exponentially with increased initial temperature, or with initial energy storage in the O:H-O bond.

1.10    Supplementary information

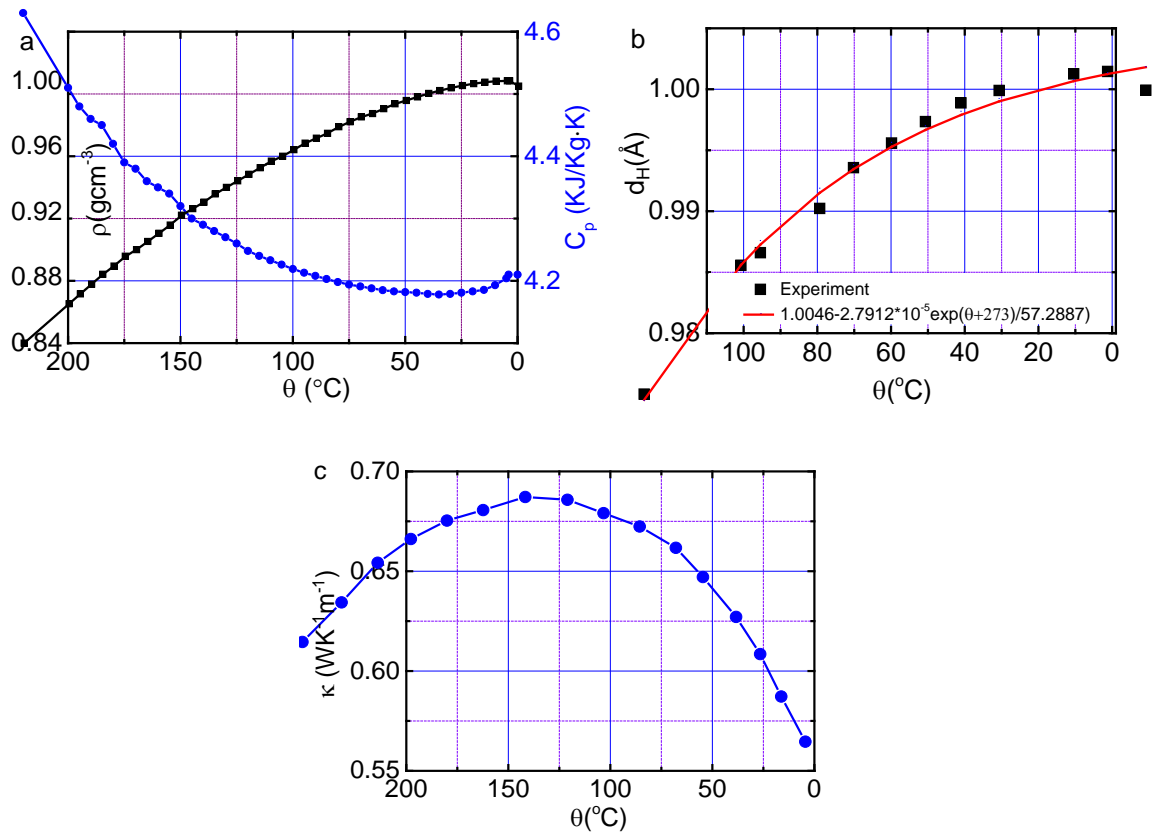

Fig A-1. Temperature dependence of (a) the ρ and C$_p$; (b) $d_H(\theta)$ (the solid line fitted to scattered data, derived using the equation: $d_{O\text{-}O} = 2.6950\rho^{-1/3}$ and $d_L / 1.6946 = 2 \times \{1 + exp[(d_H - 1.0004)/0.2428]\}^{-1}$ [29]) (c) thermal conductivity κ. (Reprinted with permission from [41].)